\newif\ifarxiv

\arxivfalse
\arxivtrue

\documentclass{ifacconf}

\usepackage{graphicx}      % include this line if your document contains figures
\usepackage{natbib}        % required for bibliography
\usepackage{amssymb}
\usepackage{amsmath} 
\usepackage{color}
\usepackage{multirow}
\usepackage[algo2e,vlined,linesnumbered]{algorithm2e}

\newcommand{\R}{\mathbb{R}}
\newcommand{\N}{\mathbb{N}}

\newtheorem{problem}{Problem}
\newtheorem{proposition}{Proposition}
\newtheorem{theorem}{Theorem}
\newtheorem{assumption}{Assumption}
\newtheorem{example}{Example}

\graphicspath{{./Images/}}
\begin{document}
\begin{frontmatter}

\title{Reachability Analysis of Neural Networks with Uncertain Parameters} 

\author[First]{Pierre-Jean Meyer} 

\address[First]{Univ Gustave Eiffel, COSYS-ESTAS, F-59666 Villeneuve d'Ascq, France (e-mail: pierre-jean.meyer@univ-eiffel.fr)}

\begin{abstract}                % Abstract of not more than 250 words.
The literature on reachability analysis methods for neural networks currently only focuses on uncertainties on the network's inputs.
In this paper, we introduce two new approaches for the reachability analysis of neural networks with additional uncertainties on their internal parameters (weight matrices and bias vectors of each layer), which may open the field of formal methods on neural networks to new topics, such as safe training or network repair.
The first and main method that we propose relies on existing reachability analysis approach based on mixed monotonicity (initially introduced for dynamical systems).
The second proposed approach extends the ESIP (Error-based Symbolic Interval Propagation) approach which was first implemented in the verification tool Neurify, and first mentioned in the publication of the tool VeriNet.
Although the ESIP approach has been shown to often outperform the mixed-monotonicity reachability analysis in the classical case with uncertainties only on the network's inputs, we show in this paper through numerical simulations that the situation is greatly reversed (in terms of precision, computation time, memory usage, and broader applicability) when dealing with uncertainties on the weights and biases.
\end{abstract}


\end{frontmatter}
%===============================================================================

%%%%%%%%%%%%%%%%%%%%%%%%%%%%%%%%%%%%%%%%%%%%%%%%%%%%%%%%%%%%%%%%%%%%%%%%%%%%%%%%
%%%%%%%%%%%%%%%%%%%%%%%%%%%%%%%%%%%%%%%%%%%%%%%%%%%%%%%%%%%%%%%%%%%%%%%%%%%%%%%%
\section{Introduction}
\label{sec intro}
In the recent years, artificial intelligence methods have grown very rapidly and spread to numerous application fields.
Although such approaches often work well in practice, the usual statistical testing of their behavior~\citep{kim2020programmatic} becomes insufficient when dealing with safety-critical applications such as autonomous vehicles~\citep{xiang2018verification}.
In such application fields, we instead need to develop formal verification approaches to guarantee the desired safe behavior of the system.
In the case of neural networks, most formal verification tools rely on reachability analysis methods or on solving optimization problems~\citep{liu2019algorithms}, and they aim to verify safety specifications in the form of input-output conditions: check if, given a set of allowed input values, the set of all outputs reachable by the network 
\ifarxiv
(or an over-approximation of this set) 
\fi
remains within safe bounds~\citep{bak2021second}.

Note however that all the tools mentioned in~\cite{liu2019algorithms,bak2021second} focus on the safety verification of pre-trained neural networks, i.e.\ the network parameters (weight matrices, bias vectors) are assumed to be fixed and known, and they only consider uncertainties on the network's input (from the safety specification to be checked).
In contrast, in this paper we are interested in expending the reachability-based verification methods 
\ifarxiv
to a whole set of neural networks, or equivalently 
\fi
to a neural network with additional uncertainties on all its internal parameters (weight matrices and bias vectors).
Such methods would in turn allow us to connect this field of neural network verification to new topics and offer new ways to approach problems such as safe training (ensuring during training that the trained network satisfies the desired properties, see e.g.~\cite{gowal2018effectiveness,mirman2018differentiable}) and network repair (finding the smallest changes to apply to an unsafe network in order to ensure its safety, see e.g.~\cite{majd2021local,yang2022neural}).
This paper thus introduces the first necessary step in the development of such verification tools: creating new methods for the reachability analysis of neural networks with bounded inputs, weights and biases.

\paragraph*{Contributions}
We propose two new methods to compute interval over-approximations of the output set of a neural network with bounded uncertainties on its inputs, weight matrices and bias vectors.
The first approach and main contribution is based on mixed-monotonicity reachability analysis (initially introduced for the analysis of dynamical systems~\citep{meyer2021springer}).
One of the main strength of this approach is its generality since it is applicable to neural networks with any Lipschitz-continuous activation functions, unlike most other approaches in the literature which are limited to piecewise-affine~\citep{shiqi2018neurify,katz2019marabou,botoeva2020efficient,xu2021fast}, sigmoid-shaped~\citep{henriksen2020efficient,tran2020nnv,eran} or monotone increasing functions~\citep{dvijotham2018dual,raghunathan2018certified}.
The proposed algorithm applies mixed-monotonicity reachability analysis to each partial network within the main neural network, and then intersects their results to obtain tighter over-approximations than if the reachability analysis was applied only once to the whole network directly.

Since, to the best of our knowledge, other approaches solving the considered problem have not yet been proposed in the literature, we introduce a second method to offer some elements of comparison with the above mixed-monotonicity approach.
This second algorithm extends to uncertain neural networks the ESIP (Error-based Symbolic Interval Propagation) method described in~\cite{henriksen2020efficient}.
Although this ESIP approach is more limited in terms of activation functions (only piecewise-affine and sigmoid-shaped functions), this method was chosen here because it was shown in~\cite{meyer2022lcss} to be very computationally efficient in the particular case of uncertainties only on the network's input.
However in this paper, numerical simulations show that with additional uncertainties on the network parameters, the mixed-monotonicity approach outperforms the ESIP algorithm on all relevant criteria: tightness of over-approximations, computation time and memory usage.

\paragraph*{Related work}
Both methods proposed in this paper to tackle the reachability analysis problem on an uncertain neural network are generalizations of the methods presented in the particular case without uncertainties on the weights and biases of the network in~\cite{meyer2022lcss} for the mixed-monotonicity approach, and in~\cite{henriksen2020efficient} for the ESIP approach.
To the best of our knowledge, the only other publication attempting to consider a similar problem in the literature is~\cite{zuo2014non}.
On the other hand, while the authors of this work indeed consider reachability analysis of uncertain neural networks, they do it in a very different setting of neural ordinary differential equations which does not allow us to provide any theoretical or numerical comparison with our approach on discrete models of feedfoward neural networks.
As mentioned above, many existing works on safety verification of neural networks also rely on various algorithms and set representations for reachability analysis (see e.g.\ those listed in the survey paper~\cite{liu2019algorithms} or the neural network verification competition~\cite{bak2021second}).
However, all these works currently only apply their reachability methods to pre-trained neural networks, and thus without any uncertainty on the weight matrices and bias vectors as we consider in this paper.

This paper is organized as follows.
Section~\ref{sec prelim} introduces the considered neural network model and defines the reachability analysis problem.
Section~\ref{sec mm} describe the first and main contribution of this paper, solving the considered problem with mixed-monotonicity reachability analysis.
The second approach based on ESIP (Error-based Symbolic Interval Propagation) is introduced in~\ref{sec esip}.
Finally, Section~\ref{sec simu} provides numerical simulations to compare both algorithms and to highlight the advantages of the mixed-monotonicity approach.

%%%%%%%%%%%%%%%%%%%%%%%%%%%%%%%%%%%%%%%%%%%%%%%%%%%%%%%%%%%%%%%%%%%%%%%%%%%%%%%%
%%%%%%%%%%%%%%%%%%%%%%%%%%%%%%%%%%%%%%%%%%%%%%%%%%%%%%%%%%%%%%%%%%%%%%%%%%%%%%%%
\section{Problem definition}
\label{sec prelim}
Given $\underline{x},\overline{x}\in\R^n$ with $\underline{x}\leq\overline{x}$, the interval $[\underline{x},\overline{x}]\subseteq\R^n$ is the set $\{x\in\R^n~|~\forall i\in\{1,\dots,n\},~\underline{x}_i\leq x_i\leq\overline{x}_i\}$.
\ifarxiv

\else
\fi
We consider an $L$-layer feedforward neural network defined as
\begin{equation}
\label{eq ffnn}
x^l=\Phi(W^lx^{l-1}+b^l),~\forall l\in\{1,\dots,L\}
\end{equation}
with uncertain input vector $x^0\in[\underline{x^0},\overline{x^0}]\subseteq\R^{n_0}$, and uncertain weight matrix $W^l\in[\underline{W^l},\overline{W^l}]\subseteq\R^{n_l\times n_{l-1}}$ and bias vector $b^l\in[\underline{b^l},\overline{b^l}]\subseteq\R^{n_l}$ for each layer $l\in\{1,\dots,L\}$.
The function $\Phi$ is defined as the componentwise application of a scalar and Lipschitz-continuous activation function.
\ifarxiv
For simplicity of presentation, the activation function $\Phi$ is assumed to be identical for all layers.
\fi

In this paper, we are interested in the robustness of the neural network with respect to the uncertainties on its input $x^0$, weights $W^l$ and biases $b^l$.
Since the output set of the network cannot be computed exactly due to the nonlinearities in the activation function $\Phi$, we use a simpler set representation (mutli-dimensional interval) to over-approximate this output set.
Relying on such over-approximations ensures that any safety property satisfied on the computed interval is guaranteed to also be satisfied on the real output set of the neural network.
This reachability analysis problem is formalized as follows.
\begin{problem}
\label{pb definition}
Given the $L$-layer neural network~\eqref{eq ffnn} and the uncertainty sets $[\underline{x^0},\overline{x^0}]\subseteq\R^{n_0}$, $[\underline{W^l},\overline{W^l}]\subseteq\R^{n_l\times n_{l-1}}$ and $[\underline{b^l},\overline{b^l}]\subseteq\R^{n_l}$ for all $l\in\{1,\dots,L\}$, find an interval $[\underline{x^L},\overline{x^L}]\subseteq\R^{n_L}$ over-approximating the output set of \eqref{eq ffnn}:
$$
\left\{
x^L\text{ in }\eqref{eq ffnn}
\left|
\begin{matrix}
x^0\in[\underline{x^0},\overline{x^0}],
W^l\in[\underline{W^l},\overline{W^l}], \\
b^l\in[\underline{b^l},\overline{b^l}],
\forall l\in\{1,\dots,L\}
\end{matrix}
\right.
\right\}
\subseteq[\underline{x^L},\overline{x^L}].
$$
\end{problem}

\ifarxiv
The secondary goal is to find over-approximations that are as close to the real output set as possible.
\fi
In this paper, we introduce two new approaches addressing this reachability analysis problem of neural networks with uncertain parameters, which has not been explored in the literature yet.
The first and main contribution in Section~\ref{sec mm} is based on mixed-monotonicity reachability analysis.
The second proposed approach in Section~\ref{sec esip} relies on Error-based Symbolic Interval Propagation (ESIP).

%%%%%%%%%%%%%%%%%%%%%%%%%%%%%%%%%%%%%%%%%%%%%%%%%%%%%%%%%%%%%%%%%%%%%%%%%%%%%%%%
%%%%%%%%%%%%%%%%%%%%%%%%%%%%%%%%%%%%%%%%%%%%%%%%%%%%%%%%%%%%%%%%%%%%%%%%%%%%%%%%
\section{Mixed monotonicity}
\label{sec mm}

%%%%%%%%%%%%%%%%%%%%%%%%%%%%%%%%%%%%%%%%%%%%%%%%%%%%%%%%%%%%%%%%%%%%%%%%%%%%%%%%
\subsection{Mixed-monotonicity reachability analysis}
\label{sub mm reach}
We first introduce the reachability analysis method for a general static function $y=f(x)$, which will then be applied multiple times to the various partial networks within \eqref{eq ffnn} in the following sections.
This result is a straightforward generalization to static functions $y=f(x)$ of the reachability analysis approach for discrete-time system $x^+=f(x)$ proposed in~\cite{meyer2021springer}.
It relies on the boundedness assumption of the derivative (also called Jacobian matrix in the paper) of function $f$, which is satisfied by any Lipschitz-continuous function.

\begin{proposition}
\label{prop algebraic MM}
Consider the function $y=f(x)$ with output $y\in\R^{n_y}$ and bounded input $x\in[\underline{x},\overline{x}]\subseteq\R^{n_x}$.
Assume that its derivative $f'$ is bounded: for all $x\in[\underline{x},\overline{x}]$, $f'(x)\in[\underline{J},\overline{J}]\subseteq\R^{n_y\times n_x}$; and denote as $J^*$ the center of these derivative bounds.
For each output dimension $i\in\{1,\dots,n_y\}$, define input vectors $\underline{\xi^i},\overline{\xi^i}\in\R^{n_x}$ and row vector $\alpha^i\in\R^{1\times n_x}$ such that for all $j\in\{1,\dots,n_x\}$,
$$(\underline{\xi^i}_j,\overline{\xi^i}_j,\alpha^i_j)=
\begin{cases}
(\overline{x}_j,\underline{x}_j,\max(0,\overline{J}_{ij}))&\text{if }J^*_{ij}<0,\\
(\underline{x}_j,\overline{x}_j,\min(0,\underline{J}_{ij}))&\text{if }J^*_{ij}\geq0.
\end{cases}$$
Then for all $x\in[\underline{x},\overline{x}]$ and $i\in\{1,\dots,n_y\}$, we have:
$$f_i(x)\in\left[f_i(\underline{\xi^i})-\alpha^i(\underline{\xi^i}-\overline{\xi^i}),\quad f_i(\overline{\xi^i})+\alpha^i(\underline{\xi^i}-\overline{\xi^i})\right].$$
\end{proposition}

\ifarxiv
Intuitively, the output bounds are obtained by computing for each output dimension the images for two diagonally opposite vertices of the input interval, then expanding these bounds with an error term when the bounds on the derivative $f'$ spans both negative and positive values.
Proposition~\ref{prop algebraic MM} can thus provide an interval over-approximation of the output set of any function as long as bounds on the derivative $f'$ are known.
Obtaining such bounds for a neural network is made possible by computing local bounds on the derivative of its activation functions, as detailed in Section~\ref{sub mm af bounds}.
\fi

%%%%%%%%%%%%%%%%%%%%%%%%%%%%%%%%%%%%%%%%%%%%%%%%%%%%%%%%%%%%%%%%%%%%%%%%%%%%%%%%
\subsection{Local bounds of activation functions}
\label{sub mm af bounds}
Proposition~\ref{prop algebraic MM} and the main algorithm in Section~\ref{sub mm algo} are applicable to neural networks with any Lipschitz-continuous activation function $\Phi$.
This is indeed a sufficient condition for the derivative of the whole network description \eqref{eq ffnn} to be bounded.
On the other hand, knowing the values of these derivative bounds is required to apply Proposition~\ref{prop algebraic MM} to the neural network.
To avoid asking users of this method to compute themselves the derivative bounds of their neural network, we restrict our framework to a subset of Lipschitz-continuous activation functions for which we provide a method to automatically define local bounding functions for the derivative of a given activation function.

\begin{assumption}
\label{assum af}
Let $\R_\infty=\R\cup\{-\infty,+\infty\}$ and consider a scalar activation function $\Phi$ whose derivative is defined as $\Phi':\R_\infty\rightarrow\R_\infty$, and where $\Phi'(x)\in\{-\infty,+\infty\}$ only if $x\in\{-\infty,+\infty\}$.
The global $\arg\min$ and $\arg\max$ $\underline{z},\overline{z}\in\R_\infty$ of $\Phi'$ are known, and $\Phi'$ is a $3$-piece piecewise-monotone function as follows:
\begin{itemize}
\item non-increasing on $(-\infty,\underline{z}]$ until reaching its global minimum $\min_{x\in\R_\infty}\Phi'(x)=\Phi'(\underline{z})$;
\item non-decreasing on $[\underline{z},\overline{z}]$ until reaching its global maximum $\max_{x\in\R_\infty}\Phi'(x)=\Phi'(\overline{z})$; 
\item and non-increasing on $[\overline{z},+\infty)$.
\end{itemize}
When $\underline{z}=-\infty$ (resp.\ $\overline{z}=+\infty$), the first (resp.\ last) monotone segment is squeezed into a singleton at infinity and can thus be ignored.
\end{assumption}

While the formulation of this assumption may seem restrictive (compared to the initial assumption of taking any Lipschitz-continuous activation function), it should be noted that the large majority of activation functions in the literature indeed have a derivative as described in Assumption~\ref{assum af}, including all the less common non-monotone activation functions reviewed or introduced in~\cite{zhu2021pflu}.\footnote{More details and examples on activation functions satisfying Assumption~\ref{assum af} are available in~\cite{meyer2022lcss} where this assumption was first introduced.}
Therefore, the mixed-monotonicity approach proposed in Section~\ref{sub mm algo} has a much broader applicability than most neural network verification tools in the literature, which are most often restricted to ReLU and piecewise-affine activation functions~\citep{shiqi2018neurify,katz2019marabou,botoeva2020efficient,xu2021fast}, occasionally able to consider sigmoid-shaped functions~\citep{henriksen2020efficient,tran2020nnv,eran}, and very rarely dealing with general monotone activation functions~\citep{dvijotham2018dual,raghunathan2018certified}.

\begin{proposition}
\label{prop local bounds}
Given an activation function $\Phi$ satisfying Assumption~\ref{assum af} and a bounded input domain $[\underline{x},\overline{x}]\in\R$, the local bounds of the derivative $\Phi'$ on $[\underline{x},\overline{x}]$ are given by:
\begin{gather*}
\label{eq local bounds}
\min_{x\in[\underline{x},\overline{x}]}\Phi'(x)=
\begin{cases}
\Phi'(\underline{z}) &\text{if }\underline{z}\in[\underline{x},\overline{x}],\\
\min(\Phi'(\underline{x}),\Phi'(\overline{x}))&\text{otherwise},
\end{cases}
\\
\max_{x\in[\underline{x},\overline{x}]}\Phi'(x)=
\begin{cases}
\Phi'(\overline{z}) &\text{if }\overline{z}\in[\underline{x},\overline{x}],\\
\max(\Phi'(\underline{x}),\Phi'(\overline{x}))&\text{otherwise}.
\end{cases}
\end{gather*}
\end{proposition}

\begin{figure}[tbp]
\centering
\includegraphics[width=\columnwidth]{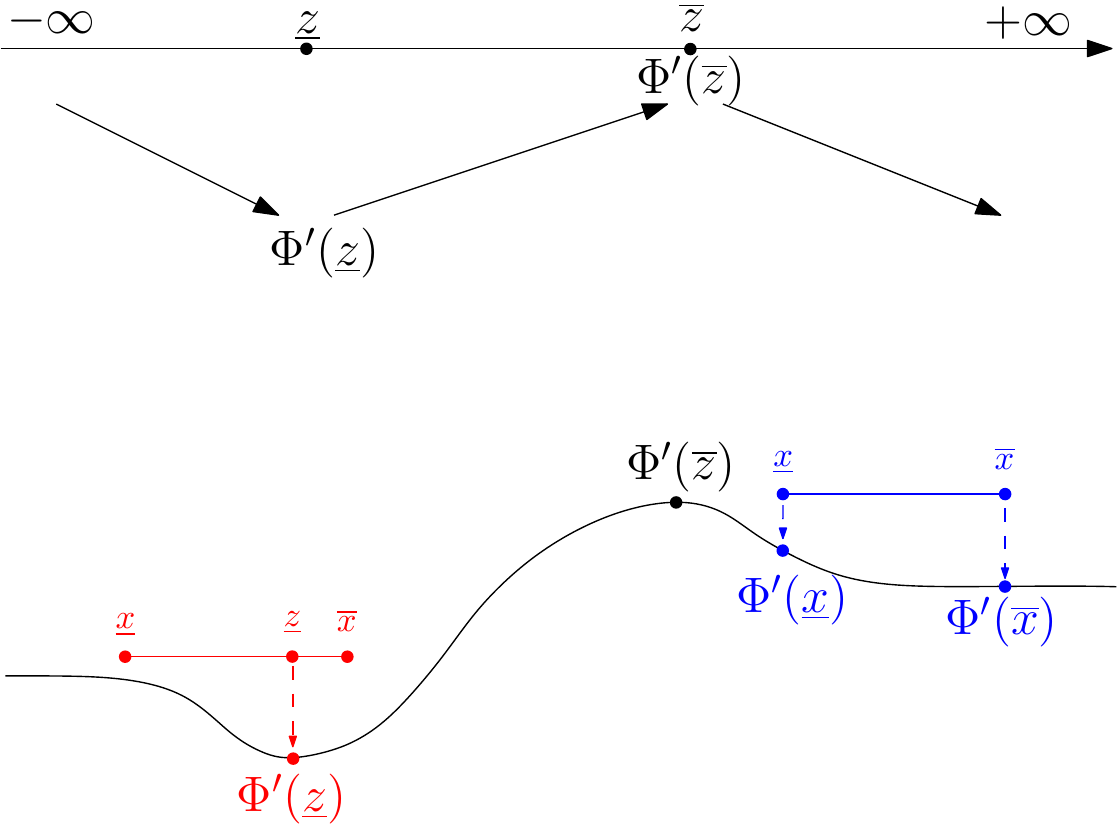}
\caption{\emph{Top}: General shape for the activation function derivative according to Assumption~\ref{assum af}. \emph{Bottom}: Two examples for the computation of the local lower bound of $\Phi'$ depending on whether its global $\arg\min$ $\underline{z}$ is contained in the input interval $[\underline{x},\overline{x}]$ (in red) or not (in blue).}
\label{fig af bounds}
\end{figure}

In short, as long as the user provides $\Phi'$ and its global $\arg\min$ and $\arg\max$ ($\underline{z}$ and $\overline{z}$), Proposition~\ref{prop local bounds} returns a local bounding function of $\Phi'$.
An illustration of Assumption~\ref{assum af} and Proposition~\ref{prop local bounds} (for the lower bound of $\Phi'$) is provided in Fig.~\ref{fig af bounds}.
The local bounding of $\Phi'$ in Proposition~\ref{prop local bounds} is used in Section~\ref{sub mm algo} for the computation of bounds on the Jacobian matrix of the neural network, which is required to apply the mixed-monotonicity reachability result from Proposition~\ref{prop algebraic MM}.

%%%%%%%%%%%%%%%%%%%%%%%%%%%%%%%%%%%%%%%%%%%%%%%%%%%%%%%%%%%%%%%%%%%%%%%%%%%%%%%%
\subsection{Main algorithm}
\label{sub mm algo}
In this section, we propose an approach using both Propositions~\ref{prop algebraic MM} and~\ref{prop local bounds} to solve Problem~\ref{pb definition} and obtain the tightest possible interval over-approximation of the neural network output set that can be computed when applying mixed-monotonicity reachability analysis to \eqref{eq ffnn}.
The proposed algorithm is inspired by the one introduced in~\cite{meyer2022lcss} in the particular case of a pre-trained neural network with only uncertainties on its input vector $x^0$.

Although Proposition~\ref{prop algebraic MM} can be applied to any partial network (described as a subset of consecutive layers of \eqref{eq ffnn}), we do not know in advance which decomposition into partial networks yields the best results.
Algorithm~\ref{algo mm} thus proposes an efficient way to apply this mixed-monotonicity reachability analysis on all possible network decompositions, while avoiding any redundant computation.
To achieve this, we explore the layers of \eqref{eq ffnn} iteratively, and apply Proposition~\ref{prop algebraic MM} to each partial network ending at the current layer.
All the obtained interval over-approximations of this layer's output set are then intersected to obtain a significantly tighter over-approximation, which will then be used in the computations of the next layers.

These main steps are summarized in Algorithm~\ref{algo mm} and described below.
The algorithm takes as input the neural network description \eqref{eq ffnn} with an activation function $\Phi$ satisfying Assumption~\ref{assum af}, as well as all the intervals bounding the network's uncertainties: network input $x^0$ and the weight matrices $W^l$ and bias vectors $b^l$ for each layer $l\in\{1,\dots,L\}$.
For a partial network considering only layers $k$ to $l$ (with $k\leq l$) of \eqref{eq ffnn} and denoted as $\mathtt{NN}(k,l)$, we use the notations: $u(k,l)$ for the concatenated vector of all its uncertainties (the partial network's input $x^{k-1}$, and the elements of all $W^i$ and $b^i$ for $i\in\{k,\dots,l\}$); $J(k,l)$ for the derivative (or Jacobian matrix) of this partial network with respect to $u(k,l)$; and $x(k,l)$ for the output of this partial network that we want to over-approximate.

Since the Jacobian bounds are defined iteratively using products, they are initialized in line~\ref{line init} as identity matrices.
Then, for each layer $l$ (lines~\ref{line for l} to~\ref{line intersection}), we first use Proposition~\ref{prop local bounds} to compute local bounds on the activation function derivative $\Phi'$ when the pre-activation variable is the result of the affine transformation of layer $l$ (line~\ref{line af der bounds}, using interval arithmetic operators): $[\underline{W^l},\overline{W^l}]*[\underline{x^{l-1}},\overline{x^{l-1}}]+[\underline{b^l},\overline{b^l}]$.

Next, we consider independently each partial network covering from some previous layer $k\in\{1,\dots,l\}$ to the current layer $l$ (lines~\ref{line for k} to~\ref{line output oa}).
The first step in line~\ref{line jacobian bounds} is to compute bounds on the Jacobian matrix of partial network $\mathtt{NN}(k,l)$.
Using the chain rule, we know that bounds on the derivative of $\mathtt{NN}(k,l)$ with respect to all uncertainties in $u(k,l)$ are given by the product:
\begin{equation}
\label{eq algo mm jac bounds}
[\underline{J(k,l)},\overline{J(k,l)}] = [\underline{\Phi'},\overline{\Phi'}]*[\underline{G(k,l)},\overline{G(k,l)}],
\end{equation}
where $G(k,l)$ is the derivative of $\mathtt{NN}(k,l)$ without the last activation function.
The bounds on $G(k,l)$ are defined as the horizontal concatenation of:
$$[\underline{W^l},\overline{W^l}]*[\underline{J(k,l-1)},\overline{J(k,l-1)}]$$
representing the derivative with respect to all uncertainties in $\mathtt{NN}(k,l-1)$;
$$[\underline{x^{l-1}}_1,\overline{x^{l-1}}_1]*I_{n_l}, \dots, [\underline{x^{l-1}}_{n_{l-1}},\overline{x^{l-1}}_{n_{l-1}}]*I_{n_l}$$
representing the derivative with respect to the weight matrix $W^l$ of the last layer;
and $I_{n_l}$ representing the derivative with respect to the bias vector $b^l$ of the last layer.
Using the computed Jacobian bounds $[\underline{J(k,l)},\overline{J(k,l)}]$ along with the known uncertainty bounds $[\underline{u(k,l)},\overline{u(k,l)}]$, we can then apply Proposition~\ref{prop algebraic MM} to the partial network $\mathtt{NN}(k,l)$ in line~\ref{line output oa} to obtain an interval over-approximation $[\underline{x(k,l)},\overline{x(k,l)}]$ of the output set of $\mathtt{NN}(k,l)$.

The last step of the algorithm, in line~\ref{line intersection}, is to take the intersection of the interval over-approximations obtained for each partial neural network ending at layer $l$.
\ifarxiv
Algorithm~\ref{algo mm} then returns the interval $[\underline{x^L},\overline{x^L}]$ computed at the last layer $L$ of the network.
\else
Algorithm~\ref{algo mm} then returns the interval $[\underline{x^L},\overline{x^L}]$ computed at the last layer $L$ of the network, which is a solution to Problem~\ref{pb definition}.
\fi

\begin{algorithm2e}[tb]
  \SetKwFunction{AlgebraicMM}{Prop1}
  \SetKwFunction{LocalBounds}{Prop2}
  \SetKwFunction{NN}{NN}
  \KwIn{$L$-layer network \eqref{eq ffnn} with activation function $\Phi$ satisfying Assumption~\ref{assum af}, uncertainties $[\underline{x^0},\overline{x^0}]\subseteq\R^{n_0}$, $[\underline{W^l},\overline{W^l}]\subseteq\R^{n_l\times n_{l-1}}$ and $[\underline{b^l},\overline{b^l}]\subseteq\R^{n_l}$ for all $l\in\{1,\dots,L\}$}
  $\forall k,l\in\{0,\dots,L\},~\underline{J(k,l)}\leftarrow I,\overline{J(k,l)}\leftarrow I$\nllabel{line init}\\
  \For{$l\in\{1,\dots,L\}$\nllabel{line for l}}{
  $[\underline{\Phi'},\overline{\Phi'}]\leftarrow\LocalBounds(\Phi',[\underline{W^l},\overline{W^l}]*[\underline{x^{l-1}},\overline{x^{l-1}}]+[\underline{b^l},\overline{b^l}])$\nllabel{line af der bounds}\\
    \For{$k\in\{1,\dots,l\}$\nllabel{line for k}}{
    Compute $[\underline{J(k,l)},\overline{J(k,l)}]$ as in \eqref{eq algo mm jac bounds}\nllabel{line jacobian bounds}\\
    $[\underline{x(k,l)},\overline{x(k,l)}]\leftarrow\AlgebraicMM(\NN(k,l),[\underline{u(k,l)},\overline{u(k,l)}],[\underline{J(k,l)},\overline{J(k,l)}])$\nllabel{line output oa}
    }
    $[\underline{x^l},\overline{x^l}]\leftarrow[\underline{x(1,l)},\overline{x(1,l)}]\cap\dots\cap[\underline{x(l,l)},\overline{x(l,l)}]$\nllabel{line intersection}
  }
  \KwOut{Over-approximation $[\underline{x^L},\overline{x^L}]$ of the 
  \ifarxiv network \fi
  output}
\caption{Mixed-monotonicity reachability analysis of an uncertain feedforward neural network.\label{algo mm}}
\end{algorithm2e}

\ifarxiv
\begin{theorem}
\label{th reachability}
The interval $[\underline{x^L},\overline{x^L}]$ returned by Algorithm~\ref{algo mm} is a solution to Problem~\ref{eq ffnn}.
\end{theorem}
\begin{pf}
Proposition~\ref{prop algebraic MM} is guaranteed in~\cite{meyer2021springer} to return an interval over-approximation for the output set of the considered system.
The theorem statement can then be proved by induction.
For layer $1$, Algorithm~\ref{algo mm} computes $[\underline{x^1},\overline{x^1}]=[\underline{x(1,1)},\overline{x(1,1)}]$, which is indeed an over-approximation of the output set of layer $1$ under all uncertainties on $x^0$, $W^1$ and $b^1$.
Next, assuming that intervals $[\underline{x^1},\overline{x^1}]$, \dots, $[\underline{x^{l-1}},\overline{x^{l-1}}]$ over-approximate the output sets of layers $1$ to $l-1$ respectively, then according to Proposition~\ref{prop algebraic MM} each interval $[\underline{x(k,l)},\overline{x(k,l)}]$ over-approximates the output set of layer $l$.
If the true output set of layer $l$ is included in each of these intervals, it is then included in their intersection $[\underline{x^l},\overline{x^l}]$.
\hfill$\blacksquare$
\end{pf}

The proof of Theorem~\ref{th reachability} thus primarily relies on the fact that the intersection operator preserves the soundness of over-approximations.
Another benefit of the use of intersections comes into play in the fact that Algorithm~\ref{algo mm} proposes an exhaustive exploration of all possible decomposition of \eqref{eq ffnn} into partial networks.
Indeed, this ensures that the final result of Algorithm~\ref{algo mm} is at least as tight as the one that would be obtained from using Proposition~\ref{prop algebraic MM} on any specific decomposition into partial networks.
And in practice, the result of Algorithm~\ref{algo mm} is often strictly tighter than any other result from specific decomposition, due to the fact that the intersection of multiple intervals is strictly smaller than each individual interval in most cases (except in the case where one is included in all others, in which case the intersection is equal to this smallest interval).
\fi

%\red{
%\begin{itemize}
%\item Include some complexity discussion (polynomial in L) ? Or a small paragraph about strneght/weakness ? (Or in separate section for comparison of both methods ?)
%\end{itemize}
%}

%%%%%%%%%%%%%%%%%%%%%%%%%%%%%%%%%%%%%%%%%%%%%%%%%%%%%%%%%%%%%%%%%%%%%%%%%%%%%%%%
%%%%%%%%%%%%%%%%%%%%%%%%%%%%%%%%%%%%%%%%%%%%%%%%%%%%%%%%%%%%%%%%%%%%%%%%%%%%%%%%
\section{Error-based Symbolic Interval Propagation}
\label{sec esip}
Although the primary contribution of this paper is the mixed-monotonicity approach presented in Section~\ref{sec mm}, numerically evaluating its performances is of great importance as for any computational method on neural networks.
Since we were not able to find any relevant and comparable approaches in the literature, we developed a second method solving Problem~\ref{pb definition}, to be able to provide numerical comparisons between them in Section~\ref{sec simu}.

The approach proposed in this section relies on symbolic interval propagation, which has been used in several neural network verification tools such as ReluVal~\citep{shiqi2018reluval}, Neurify~\citep{shiqi2018neurify} and VeriNet~\citep{henriksen2020efficient}.
The core idea is to create bounding functions depending linearly in the network input, and propagating these linear functions iteratively through the layers of the network (i.e.\ through both the layer's affine transformation, and the linear relaxations of the nonlinear activation functions~\citep{salman2019convex}).
This ensures that the dependency to the network's input is preserved during the propagation of these bounding functions through the network, which results in significantly tighter reachable set over-approximations compared to naive interval bound propagation approaches (see e.g.~\cite{xiang2020reachable}) where this input dependency is lost at each layer.

In this paper, we are interested in a particular variation called ESIP (Error-based Symbolic Interval Propagation), which was first introduced (but not published) in the implementation of the tool Neurify~\citep{shiqi2018neurify} and first published in the paper of the tool VeriNet~\citep{henriksen2020efficient}.
Unlike the classical approach designing and propagating two linear functions (for the lower and upper bounds, respectively), ESIP relies on a single linear function alongside an error matrix.
The symbolic equation represents the behavior of the network if the nonlinear activation function of each node is replaced by its lower linear relaxation.
The error matrix accounts for deviations from this symbolic equation induced by nodes operating at the upper linear relaxation of their activation function.
This particular approach is chosen here because it was shown in~\cite{meyer2022lcss} to be very computationally efficient, with a low and constant computation time per node in the network, while other methods had a computation time per node that grew with the size of the network.

Compared to the ESIP implementation in~\cite{henriksen2020efficient} in the case where the neural network only has uncertainties on its input, we propose here an extension of this approach to also handle uncertainties on the weight matrices and bias vectors and thus to be able to solve Problem~\ref{pb definition}.
This extension is summarized in Algorithm~\ref{algo esip} and detailed below.
Due to space limitations and since this new approach is primarily introduced for comparison with our main contribution in Section~\ref{sec mm}, the formal proof that Algorithm~\ref{algo esip} solves Problem~\ref{pb definition} is left for future work.
We refer the reader to~\cite{henriksen2020efficient} for more theoretical details in the case of uncertainties only on the network input $x^0$.

In line~\ref{esip line init}, we initialize the uncertainty vector $u^0$ to the input $x^0$, the symbolic equation $S^0$ to the identity function, and the error $[\underline{E},\overline{E}]$ to and empty interval matrix.
Then, for each layer $l$ we first propagate these variables through the affine transformation $x\rightarrow W^l*x+b^l$, by appending all elements of $W^l$ and $b^l$ at the end of the previous uncertainty vector $u^{l-1}$ (line~\ref{esip line affine uncertainty}), updating the symbolic equation with respect to the new uncertainties $W^l$ and $b^l$ introduced in this layer (line~\ref{esip line affine symbolic}), and multiplying the error by the bounds on $W^l$ (line~\ref{esip line affine error}).

Next, propagating through the layer's activation function is done individually for each node $i$ of the layer (line~\ref{esip line for i}).
We first compute concrete bounds of the pre-activation variable (line~\ref{esip line pre activation}) by evaluating the symbolic equation $S^l_i$ on the current uncertainty bounds $[\underline{u^l},\overline{u^l}]$, and then adding all negative error terms to the lower bound and all positive errors to the upper bound.
These pre-activation bounds can then be used in line~\ref{esip line af relaxation} to compute a linear relaxation of the activation function, i.e.\ two linear functions $\underline{r}$ and $\overline{r}$ such that $\underline{r}(x)\leq \Phi(x)\leq\overline{r}(x)$ for all $x\in[\underline{x^l}_i,\overline{x^l}_i]$.
More details on how to compute such linear relaxations can be found e.g.\ in~\cite{xu2021fast} for ReLU functions and in~\cite{henriksen2020efficient} for sigmoid-shaped functions.
We then propagate the symbolic equation through the lower relaxation $\underline{r}$ (line~\ref{esip line af symbolic}) and compute the maximal error between the relaxation bounds over the pre-activation range $[\underline{x^l}_i,\overline{x^l}_i]$ (line~\ref{esip line af error scalar}).
These new error terms are appended at the end of both bounds of the error (line~\ref{esip line af error matrix}).

For the final layer $L$, the propagation of the symbolic equation and error through the activation function (lines~\ref{esip line af relaxation}-\ref{esip line af error matrix}) can be skipped since the interval over-approximation of the output set can be simply computed by propagating the pre-activation bounds (from line~\ref{esip line pre activation}) through the activation function.
Note that in line~\ref{esip line final oa}, we obtain these bounds by applying $\Phi$ directly to the lower and upper bounds because this class of approaches relying on linear relaxations are currently limited in the literature (either in their theory or their implementation) to monotone increasing activation functions~\citep{shiqi2018neurify,henriksen2020efficient,zhang2018efficient}.

\begin{algorithm2e}[tbh]
  \SetKwFunction{NN}{NN}
  \KwIn{$L$-layer network \eqref{eq ffnn}, uncertainties $[\underline{x^0},\overline{x^0}]\subseteq\R^{n_0}$, $[\underline{W^l},\overline{W^l}]\subseteq\R^{n_l\times n_{l-1}}$ and $[\underline{b^l},\overline{b^l}]\subseteq\R^{n_l}$ for all $l\in\{1,\dots,L\}$}
  $u^0\leftarrow x^0$, $S^0(u^0)\leftarrow x^0$, $\underline{E} \leftarrow []$, $\overline{E} \leftarrow []$\nllabel{esip line init}\\
  \For{$l\in\{1,\dots,L\}$\nllabel{esip line for l}}{
  \tcc{Affine transformation}
  $u^l\leftarrow[u^{l-1};W^l(:);b^l]$\nllabel{esip line affine uncertainty}\\
  $S^l(u^l)\leftarrow W^l*S^{l-1}(u^{l-1})+b^l$\nllabel{esip line affine symbolic}\\
  $[\underline{E},\overline{E}]\leftarrow[\underline{W^l},\overline{W^l}]*[\underline{E},\overline{E}]$\nllabel{esip line affine error}\\
    \For{$i\in\{1,\dots,n_l\}$\nllabel{esip line for i}}{
    \tcc{Pre-activation bounds}
        $[\underline{x^l}_i,\overline{x^l}_i]\leftarrow S^l_i([\underline{u^l},\overline{u^l}])+[\sum_{j|\underline{E}(i,j)<0}\underline{E}(i,j),\sum_{j|\overline{E}(i,j)>0}\overline{E}(i,j)]$\nllabel{esip line pre activation}\\
    \tcc{Activation function}
%    $\underline{r}(x)\leftarrow\underline{c}*x+\underline{d}$, $\overline{r}(x)\leftarrow\overline{c}*x+\overline{d}$\nllabel{esip line af relaxation}\\
    find $\underline{r},\overline{r}~|~\underline{r}(x)\leq \Phi(x)\leq\overline{r}(x),\forall x\in[\underline{x^l}_i,\overline{x^l}_i]$\nllabel{esip line af relaxation}\\
    $S^l_i(u^l)\leftarrow \underline{r}(S^l_i(u^l))$\nllabel{esip line af symbolic}\\
%    $e_i\leftarrow \max(\underline{x^l}_i*(\overline{c}-\underline{c})+\overline{d}-\underline{d},\overline{x^l}_i*(\overline{c}-\underline{c})+\overline{d}-\underline{d})$\\
    $e_i\leftarrow \max(\overline{r}(\underline{x^l}_i)-\underline{r}(\underline{x^l}_i),\overline{r}(\overline{x^l}_i)-\underline{r}(\overline{x^l}_i))$\nllabel{esip line af error scalar}\\
    }
        $\underline{E}\leftarrow[\underline{E},diag(e)]$, $\overline{E}\leftarrow[\overline{E},diag(e)]$\nllabel{esip line af error matrix}\\
  }
  $[\underline{x^L},\overline{x^L}]\leftarrow [\Phi(\underline{x^L}),\Phi(\overline{x^L})]$\nllabel{esip line final oa}\\
  \KwOut{Over-approximation $[\underline{x^L},\overline{x^L}]$ of the 
  \ifarxiv network \fi
  output}
\caption{ESIP reachability analysis of an uncertain feedforward neural network.\label{algo esip}}
\end{algorithm2e}

Algorithm~\ref{algo esip} is very similar to the ESIP approach described in~\cite{henriksen2020efficient} in the particular case without weight and bias uncertainty.
The main differences are that in our Algorithm~\ref{algo esip}, the dimension of uncertainty vector $u^l$ grows at each layer, and the error needs to be described as an interval matrix (due to the product with uncertain weight $W^l$ in line~\ref{esip line affine error}) instead of a simple matrix in~\cite{henriksen2020efficient}.

In terms of implementation of the algorithm however, there is a much more significant difference with~\cite{henriksen2020efficient}: the symbolic equation which was a linear function in the network intput $x^0$ in~\cite{henriksen2020efficient} is now a multi-linear function in the uncertainty vector $u^l$.
This introduces a significantly higher complexity in terms of implementation and memory usage.
Indeed, in~\cite{henriksen2020efficient} the symbolic equation of layer $l$ could be simply defined as an $n_l\times(n_0+1)$ matrix, where for each of the $n_l$ output nodes of this layer we only need to store $n_0$ values for the factors multiplying the terms of $x^0$, and the final value for the constant term of the linear equation.
On the other hand, for the multi-linear function $S^l$ in Algorithm~\ref{algo esip}, we would similarly need to store one factor for each multi-linear term appearing in the equation.
\ifarxiv
The creation of such huge matrices thus limits the application of this ESIP approach to very shallow and narrow neural networks, as illustrated in Example~\ref{ex esip} below.
\else
The creation of such huge matrices thus limits the application of this ESIP approach to very shallow and narrow neural networks.
\fi

\ifarxiv
\begin{example}
\label{ex esip}
The initial symbolic equation $S^0$ has dimensions $n_{S^0}=n_0\times (n_0+1)$.
Next if $S^{l-1}$ is stored as an $n_{l-1}\times n_{S^{l-1}}$ matrix, then the affine transformation at layer $l$ ($S^l(u^l)= W^l*S^{l-1}(u^{l-1})+b^l$), implies that the new symbolic equation $S^l$ has $n_l*n_{l-1}*n_{S^{l-1}}$ columns for the multi-linear and linear terms in $W^l*S^{l-1}(u^{l-1})$, followed by $n_l$ columns for the linear terms in $b^l$, and a final column for the constant of the symbolic equation (which becomes a non-zero value only after its propagation through the activation function).
Therefore, $S^l$ is stored as an $n_l\times n_{S^l}$ matrix, with $n_{S^l}=n_l*n_{l-1}*n_{S^{l-1}} + n_l + 1$.

%n0+1
%n1n0²+n1n0+n1+1
%n2n1²n0²+n2n1²n0+n2n1²+n2n1+n2+1
%n3n2²n1²n0²+n3n2²n1²n0+n3n2²n1²+n3n2²n1+n3n2²+n3n2+n3+1\item

For simplicity, assume that all layers have the same width: $\exists n\in\N~|~\forall l\in\{0,\dots,L\},~n_l=n$.
Then we can prove by induction that the width of the matrix for $S^l$ is $n_{S^l}=\sum_{i=0}^{2l+1}n^{i}$.
We can thus conclude that the final symbolic equation $S^L$ is of dimensions:
$$n\times\left(\frac{1-n^{2L+2}}{1-n}\right).$$
Therefore, the complexity of Algorithm~\ref{algo esip} is exponential in the depth $L$ of the network, and polynomial in its width $n$.

In Matlab, matrices cannot contain more than $2^{48}-1\approx 2.8*10^{14}$ elements.
To illustrate the high memory usage of this approach, we show in Table~\ref{table esip} the maximum width $n$ of an $L$-layer network for the symbolic equation $S^L$ to remain within this Matlab constraint.

\begin{table}[htb]
\centering
\begin{tabular}{|c|cccccc|} 
\hline
{\bf Depth $L$} & $1$ & $2$ & $3$ & $4$ & $5$ & $6$\\
\hline
{\bf Max allowed width $n$} & $4095$ & $255$ & $63$ & $27$ & $15$ & $10$\\
\hline
\end{tabular}
\caption{Maximum width $n$ of the network for the symbolic equation $S^L$ to be storable in Matlab.\label{table esip}}
\end{table}

Note however that this constraint on the matrix dimension is never actually reached in practice, since we would first reach another limitation related to the actual weight of this matrix compared to the available RAM.
Taking the same conditions that will be considered in the numerical example of Section~\ref{sec simu} with a network of depth $L=3$ and width $n=20$, the matrix storing the symbolic equation would weigh up to $215$ GB (counting $8$ bytes per element in the matrix).
This is significantly higher than the available RAM on most computers, which will result in a crash of Matlab when attempting to create such matrix.
\hfill$\blacktriangle$
\end{example}
\fi

%\red{
%\begin{itemize}
%\item Change the memory example into a Proposition of the size/complexity with proper proof ? (And keep example for numerical values ?)
%\item If I have space for a formal theorem/proof, highlight the fact that monotonicity of AF is required (or rather, user-provided linear relaxation method). Proof might be doable by induction, with 2 steps for affine + AF ?
%\end{itemize}
%}

%%%%%%%%%%%%%%%%%%%%%%%%%%%%%%%%%%%%%%%%%%%%%%%%%%%%%%%%%%%%%%%%%%%%%%%%%%%%%%%%
%%%%%%%%%%%%%%%%%%%%%%%%%%%%%%%%%%%%%%%%%%%%%%%%%%%%%%%%%%%%%%%%%%%%%%%%%%%%%%%%
\section{Numerical examples}
\label{sec simu}
In this section, we provide a numerical comparison of Algorithms~\ref{algo mm} and~\ref{algo esip} on a set of randomly generated neural networks with various dimensions and activation functions, and we highlight the better performances of the mixed-monotonicity approach from Section~\ref{sec mm} with respect to most criteria (generality, tightness, computation time, memory usage).
Both algorithms are implemented in Matlab 2021b and run on a laptop with $1.80$GHz processor and $16$GB of RAM.

In our first numerical experiment, we consider neural networks as in \eqref{eq ffnn} with increasing depth $L$ and a fixed uniform width $n$ for all input, hidden and output layers (i.e.\ $n_l=n$ for all $l\in\{0,\dots,L\}$).
Since we already know from Example~\ref{ex esip} that the ESIP approach will struggle in terms of complexity and memory usage, we focus this comparison on narrow networks with $n=20$ nodes per layer.
All uncertainty variables (input, weight matrices, bias vectors) are assigned randomly generated bounds within $[-1,1]$.
The simulation are run a total of $N=10$ times, each with different random uncertainty bounds, and the obtained results in terms of width of the interval over-approximation, computation time and memory usage are averaged over this number of runs.
Since the original ESIP implementation in VeriNet~\citep{henriksen2020efficient} is limited to piecewise-affine or sigmoid-shaped activation functions, we focus this first comparison on the most popular activation function: Rectified Linear Unit (ReLU), which is the piecewise-affine function $\Phi(x)=\max(0,x)$.

In Table~\ref{table simu relu} are summarized the obtained results for both Algorithm~\ref{algo mm} using mixed-monotonicity and Algorithm~\ref{algo esip} using the ESIP approach.
In terms of the width of the computed interval over-approximations, we first notice that both algorithms return identical results for shallow networks ($L=1$), but that the mixed-monotonicity approach always generates tighter intervals for networks with hidden layers.
In terms of complexity, the superiority of the mixed-monotonicity approach is striking, since the computation times are on average $12$ times faster than ESIP with one layer, and up to $7000$ times faster with two layers.
Similarly the memory usage is on average $1.4$ times smaller than ESIP with one layer, and $176$ times smaller with two layers.
As predicted in Example~\ref{ex esip}, as soon as we add a third layer, the ESIP approach attempts to create a matrix much larger than the available $16$ GB of RAM (even despite the use of sparse matrices), which causes Matlab to crash.
On the other hand, we can see that the mixed-monotonicity approach from Algorithm~\ref{algo mm} still behaves well in terms of complexity (time and memory) for deeper networks, although the conservativeness of the over-approximation naturally increases with the size of the network.

\begin{table}[tbp]
\centering
\begin{tabular}{c c | c c} 
& & {\bf Mixed-monotonicity} & {\bf ESIP}\\\hline
\multirow{3}{*}{$L=1$} & 
{\bf Width} & $21.2$ & $21.2$\\
&{\bf Time} (s) & $0.067$ & $0.79$\\
&{\bf Memory} (MB) & $0.16$ & $0.22$\\\hline
\multirow{3}{*}{$L=2$} & 
{\bf Width} & $201$ & $263$\\
&{\bf Time} (s) & $0.25$ & $368$\\
&{\bf Memory} (MB) & $0.46$ & $81$\\\hline
\multirow{3}{*}{$L=3$} & 
{\bf Width} & $2144$ & $-$\\
&{\bf Time} (s) & $0.64$ & $-$\\
&{\bf Memory} (MB) & $0.90$ & $>16000$\\\hline
\multirow{3}{*}{$L=4$} & 
{\bf Width} & $21796$ & $-$\\
&{\bf Time} (s) & $1.4$ & $-$\\
&{\bf Memory} (MB) & $1.5$ & $-$\\\hline
\multirow{3}{*}{$L=5$} & 
{\bf Width} & $284625$ & $-$\\
&{\bf Time} (s) & $2.5$ & $-$\\
&{\bf Memory} (MB) & $2.2$ & $-$\\\hline
\multirow{3}{*}{$L=6$} & 
{\bf Width} & $2866970$ & $-$\\
&{\bf Time} (s) & $4.2$ & $-$\\
&{\bf Memory} (MB) & $3.1$ & $-$\\\hline
\multirow{3}{*}{$L=10$} & 
{\bf Width} & $44421135944$ & $-$\\
&{\bf Time} (s) & $18$ & $-$\\
&{\bf Memory} (MB) & $7.9$ & $-$\\\hline
\end{tabular}
\caption{Average width $\|\overline{x^L}-\underline{x^L}\|_2$ of the interval over-approximation, computation time (in seconds) and memory usage (in megabytes) for both algorithms over $10$ ReLU networks.}
\label{table simu relu}
\end{table}

\bigskip
The second set of numerical experiments is run only by the mixed-monotonicity approach in Algorithm~\ref{algo mm} and focuses on its performances while dealing with the main two limitations of the ESIP approach that we could not explore in the comparison of Table~\ref{table simu relu}: larger networks and more uncommon activation functions.
Indeed, the ESIP approach is not only limited by its complexity, but as mentioned in Section~\ref{sec esip}, Algorithm~\ref{algo esip} and its original version in VeriNet~\citep{henriksen2020efficient} also cannot handle non-monotone activation functions.
Here, we thus consider neural networks with the Sigmoid Linear Unit (SiLU) activation function, which is the non-monotone function $\Phi(x)=x/(1+e^{-x})$ introduced in~\cite{ramachandran2017searching}.
This SiLU activation function satisfies Assumption~\ref{assum af} (with global $\arg\min$ and $\arg\max$ of its derivative defined as $\underline{z}=-2.3994$ and $\overline{z}=2.3994$, respectively), and it is thus natively handled by the mixed-monotonicity approach in Algorithm~\ref{algo mm}.

Tables~\ref{table simu silu time} and~\ref{table simu silu memory} report the average (over $N=10$ randomly generated uncertainty bounds as in the previous test) computation time and memory usage, respectively, when the depth $L$ of the neural network goes from $1$ to $10$ and its width $n$ from $20$ to $100$.
Although both these quantities naturally increase with the size ($L$ or $n$) of the network, we can observe that Algorithm~\ref{algo mm} could solve Problem~\ref{pb definition} on all neural networks of up to $10$ layers and $100$ neurons per layer, in less than an hour and using less than $1$ GB of RAM.
This is a significant advantage compared to Algorithm~\ref{algo esip} which took over $6$ minutes per network for a $2$-layer $20$-width network, and over hundreds of GB for a $3$-layer network.
After plotting the obtained results from Tables~\ref{table simu silu time} and~\ref{table simu silu memory} using $\log$ and various $n$-th roots to identify the growth rates, we have identified that the mixed-monotonicity approach in Algorithm~\ref{algo mm} has a polynomial complexity in $O(n^3*L^3)$ for the computation time and $O(n^3*L^2)$ for the memory.

\begin{table}[tbp]
\centering
\begin{tabular}{c | c c c c c} 
& {\bf $n=20$} & {\bf $n=40$} & {\bf $n=60$} & {\bf $n=80$} & {\bf $n=100$}\\\hline
{\bf $L=1$} & $0.068$ & $0.40$ & $1.4$ & $3.6$ & $7.7$\\
{\bf $L=2$} & $0.24$ & $2.0$ & $7.2$ & $18$ & $38$\\
{\bf $L=3$} & $0.66$ & $5.6$ & $21$ & $51$ & $108$\\
{\bf $L=4$} & $1.4$ & $12$ & $43$ & $108$ & $225$\\
{\bf $L=5$} & $2.6$ & $21$ & $76$ & $191$ & $402$\\
{\bf $L=6$} & $4.1$ & $34$ & $124$ & $312$ & $665$\\
{\bf $L=7$} & $6.2$ & $54$ & $191$ & $486$ & $1028$\\
{\bf $L=8$} & $9.0$ & $77$ & $276$ & $705$ & $1499$\\
{\bf $L=9$} & $13$ & $109$ & $389$ & $990$ & $2109$\\
{\bf $L=10$} & $17$ & $146$ & $526$ & $1333$ & $2844$\\
\end{tabular}
\caption{Average computation time (in seconds) of Algorithm~\ref{algo mm} over $10$ SiLU networks.}
\label{table simu silu time}
\end{table}

\begin{table}[tbp]
\centering
\begin{tabular}{c | c c c c c} 
& {\bf $n=20$} & {\bf $n=40$} & {\bf $n=60$} & {\bf $n=80$} & {\bf $n=100$}\\\hline
{\bf $L=1$} & $0.16$ & $1.1$ & $3.7$ & $8.6$ & $17$\\
{\bf $L=2$} & $0.46$ & $3.3$ & $11$ & $26$ & $50$\\
{\bf $L=3$} & $0.89$ & $6.6$ & $22$ & $51$ & $99$\\
{\bf $L=4$} & $1.5$ & $11$ & $36$ & $85$ & $164$\\
{\bf $L=5$} & $2.2$ & $16$ & $54$ & $127$ & $246$\\
{\bf $L=6$} & $3.0$ & $23$ & $76$ & $178$ & $345$\\
{\bf $L=7$} & $4.1$ & $31$ & $101$ & $237$ & $459$\\
{\bf $L=8$} & $5.2$ & $39$ & $130$ & $304$ & $590$\\
{\bf $L=9$} & $6.5$ & $49$ & $162$ & $380$ & $737$\\
{\bf $L=10$} & $7.9$ & $60$ & $198$ & $464$ & $901$\\
\end{tabular}
\caption{Average memory usage (in megabytes) of Algorithm~\ref{algo mm} over $10$ SiLU networks.}
\label{table simu silu memory}
\end{table}

%\red{
%\begin{itemize}
%\item Should we try to include some comparison to RS interval hull ? (need to compute sampled outputs for each separate NN !)
%\item Upload/share the code used ?
%\end{itemize}
%}

%%%%%%%%%%%%%%%%%%%%%%%%%%%%%%%%%%%%%%%%%%%%%%%%%%%%%%%%%%%%%%%%%%%%%%%%%%%%%%%%
%%%%%%%%%%%%%%%%%%%%%%%%%%%%%%%%%%%%%%%%%%%%%%%%%%%%%%%%%%%%%%%%%%%%%%%%%%%%%%%%
\section{Conclusions}
\label{sec conclu}
In this paper, we consider the reachability analysis problem for neural networks with uncertainties not only on their inputs, but also on all their weight matrices and bias vectors.
To the best of our knowledge, this problem has not yet been addressed in the literature.
We propose two approaches to tackle this problem.
The first one and our main contribution relies on a repeated call of mixed-monotonicity reachability analysis on each partial network within the main neural network.
The second approach, primarily provided to offer some elements of comparison with our first approach in this yet unexplored topic, extends to uncertain networks the ESIP (Error-based Symbolic Interval Propagation) approach from~\cite{henriksen2020efficient}.
In both the theoretical sections and the numerical simulations, we highlight the superiority of the mixed-monotonicity approach with respect to all criteria relevant to solving the considered problem.
Indeed, the algorithm is widely applicable to networks with any Lipschitz-continuous activation function, while the ESIP approach is limited to piecewise-affine and sigmoid-shaped functions.
In terms of computation time and memory usage, the mixed-monotonicity approach has only a polynomial complexity in both the depth and width of the network, while the ESIP complexity is exponential in the network's depth which limits it to only shallow networks.
Finally, on all networks where the ESIP algorithm could be run, the mixed-monotonicity approach returns tighter interval over-approximation of the output set (or equal to ESIP in the case of a single-layer network).

While most verification tools in the neural network literature currently focus on verifying a single pre-trained neural network, the work presented in this paper instead analyzes a whole family of neural networks for any weight matrices and bias vectors in their respective bounds.
This opens the door to new topics such as safe training (finding the subset of weight and bias values such that the resulting networks satisfy a given property) or network repair (finding the minimal changes to apply to a given network in order to make it satisfy a given property), which will be the main focus of our future work.

%anytime training

%%%%%%%%%%%%%%%%%%%%%%%%%%%%%%%%%%%%%%%%%%%%%%%%%%%%%%%%%%%%%%%%%%%%%%%%%%%%%%%%
%%%%%%%%%%%%%%%%%%%%%%%%%%%%%%%%%%%%%%%%%%%%%%%%%%%%%%%%%%%%%%%%%%%%%%%%%%%%%%%%

%\bibliographystyle{ifacconf}
\bibliography{2023_Meyer_IFAC23}

\end{document}